\documentclass[preprint]{elsarticle}
\usepackage{amsmath}
\usepackage{amssymb}

\begin{document}

\title{About disposition of energy levels}

\author{Evgeny Z. Liverts\corref{cor1}}
\ead{liverts@phys.huji.ac.il}

\author{Nir Nevo Dinur}

\cortext[cor1]{Corresponding author}

\address{Racah Institute of Physics, The Hebrew University, Jerusalem 91904,
Israel}

\begin{abstract}
The unique properties of central potential of the form $-\beta e^{-r}r^{\gamma}$
were studied using the recently developed critical parameter technique.
The particular cases of $\gamma=0$ and $\gamma=-1$ yield, respectively,
the exponential and Yukawa potentials widely used in the atomic, molecular
and nuclear physics. We found different behavior of the energy levels
of this potential for three different ranges of the value of $\gamma$.
For $\gamma\geq0$ it was found that the energy of bound states with
the same principal quantum number $N$ decreases with increasing angular
momentum $\ell$. The Gaussian and Woods-Saxon potentials also show
this behavior. On the contrary, for $-2\leq\gamma\leq-1$ increasing
$\ell$ gives a higher energy, resembling the Hulthen potential. However,
a potential with $-1<\gamma<0$ possesses mixed properties, which
give rise to several interesting results. For one, the order of energy
levels with different quantum numbers is not preserved when varying
the parameter $\beta$. This leads to a quantum degeneracy of the
states, and in fact, for a given value of $\gamma$
we can find the values $\beta_{thr}$ for which two energy levels
with different quantum numbers coincide. Another interesting phenomena
is the possibility, for some values of $\gamma$ in this range, for
two new energy levels with different quantum numbers to appear simultaneously
when $\beta$ reaches their common critical value.
\end{abstract}


\maketitle

\section{Introduction}

The interest in the ordering of energy levels is as old as quantum
mechanics. It can be reflected in the electron configurations of the
elements in the periodic table, and in the nuclear shell model. In
the 1980s there was a renewed interest in this subject following the
discoveries of mesons composed of the heavier quarks $c$ and $b$.
From the disposition of the energy levels of these mesons, the nature
of the quark-quark (or, rather quark-antiquark) potential can be deduced,
if one can relate the properties of the potential with the ordering
of the energy levels pertaining to it. This development led the authors
of \cite{BGM} (following \cite{GM} and \cite{FFD}) to prove that
for non-relativistic two-body systems the order of energy levels with
the same principal quantum number $N$ is controlled by the sign of
the Laplacian of the spherically symmetric potential $V(r)$,
\begin{equation}
\Delta V(r)=\frac{2}{r}\frac{dV}{dr}+\frac{d^{2}V}{dr^{2}},
\label{eq:1 Lap}
\end{equation}
where $N=n+\ell+1$, $n$ is the number of nodes in
the radial wave function and $\ell$ is the angular momentum quantum number.
Specifically, they show that under
the following two conditions the ordering of energy levels with the
same principal quantum number $N$ can be exactly determined%
\footnote{Notice that for the 2$^{nd}$ case only the weaker condition is given
here, instead of the more restrictive condition, $\Delta V(r)<0\textrm{ for all }r>0$,
which is the opposite of \eqref{eq:2+}.}:
\begin{enumerate}
\item If
\begin{equation}
\Delta V(r)>0\textrm{ for all }r>0\label{eq:2+}
\end{equation}
then\\
\begin{equation}
E_{n,\ell}>E_{n-1,\ell+1}\label{eq:3dec}
\end{equation}

\item If
\begin{equation}
\Delta V(r)<0\textrm{ for }r<r_{0}\textrm{, and }\frac{dV}{dr}<0\textrm{ for }r\geq r_{0}>0\label{eq:4-},
\end{equation}
then\\
\begin{equation}
E_{n,\ell}<E_{n-1,\ell+1}\label{eq:5inc}
\end{equation}

\end{enumerate}
These results were also generalized for the relativistic case in Ref.
\cite{GMS}, but in this letter we limit ourselves to the non-relativistic
realm.

The question arises from these results $\textendash$ what can one say about
the ordering of energy levels relating to potentials that do not fulfill
either of the conditions \eqref{eq:2+} or \eqref{eq:4-} above? In
this letter we try to answer this question for the exponential-power
(EP) potential, i.e., of the form
\begin{equation}
V_{\beta\gamma}\left(r\right)=-\beta e^{-r}r^{\gamma}~~~~~~(\beta>0,~\gamma>-2).
\label{eq:6 the EP V}
\end{equation}
This is done using the critical parameter technique (CPT) recently
introduced in \cite{LEZ}, where it was also used to study the energy
levels of a few short-range central potentials widely used in atomic, molecular and nuclear physics.
Two of these, namely the exponential and Yukawa potentials
are special cases of $V_{\beta\gamma}$ with $\gamma$ equals $0$ and $-1$, respectively.


\section{Methods}

In \cite{LEZ} the CPT was introduced as a methodology for obtaining information
on the bound solutions of the non-relativistic Schr\"{o}dinger equation with
central potentials of the form
\begin{equation}
V\left(r\right)=-V_{0}f\left(r/r_{s}\right),~~~~~(V_0>0)
\label{eq:7 V0 f(r)}
\end{equation}
where $r_s$ is the so called screening parameter, and $V_0$ represents the coupling constant.
Taking into account the scaling properties,
the solution of the relevant equation depends effectively on a single parameter
$\beta=2mgr_{s}^{2+\gamma}/\hbar^{2}$, where $m$ is the reduced mass of the considered system.
Parameter $g$, which is proportional to $V_0$, will be defined in \eqref{eq:9 V of alpha}.
The CPT consists primarily of finding at what values of $\beta=\beta_{n,\ell}$, as it is
being varied, an eigenstate of the Hamiltonian with given $\{n,\ell\}$ becomes a transition state, 
i.e., will have zero energy. Further increasing $\beta$ beyond this ``critical''
value (CP) will make this eigenstate a bound state, causing a new bound
energy level $E_{n,\ell}$ to ``appear''.


This technique allows one to answer such questions as a) How many bound
states with given $\ell$ exist for a given potential? b) What
is the maximum $\ell$ a bound state can have, given the
potential? etc.

In \cite{LEZ} it is also shown that the CPT can determine
the order in which the different energy levels {}``appear'', as
$\beta$ is increased. This led the authors to \textit{conjecture} that this
order of the energy levels is preserved when $\beta$ is further increased
beyond the values at which the levels first appeared. If this conjecture
is true, then one can very easily determine the ordering of energy
levels of every potential that can be solved using the CPT. For example,
the authors of \cite{LEZ} derive the conditions (67), regarding the
ordering of the energy levels of the potentials studied there, by
relying only on their knowledge regarding the relations between the
CP's obtained for each of these potentials. 
Our results suggest that the validity of the conjecture depends on the
behavior of the potential near the origin, as shown in
several examples below.

A few technical details are due here. All calculations were provided
by the Mathematica 8 \cite{WM} codes which realized methods described
in \cite{LEZ}. The CP's depend on the asymptotic behavior  ($r\rightarrow\infty$)
of $\chi(r)$, the solution to the corresponding Schr\"{o}dinger equation
\begin{equation}
\frac{d^{2}\chi(r)}{dr^{2}}=\left[-\beta v(r)+\frac{\ell(\ell+1)}{r^{2}}-E_{n,\ell}\left(\beta\right)\right]\chi(r)
\label{eq:8 Eq.for chi}
\end{equation}
when
\[
E_{n,\ell}\left(\beta\right)\xrightarrow[\beta\rightarrow\beta_{n,\ell}]{}0.
\]
However, the numerical integration of the above differential equation
starts from some small value of $r$. Therefore, the accuracy of the
asymptotic behavior is defined by our possibility to calculate accurately
the radial wave function (and its derivative) 
at this initial value of $r$ which should be chosen
as far as possible from the origin.
Let the leading term of a series expansion for the potential under
consideration has the form:
\begin{equation}
v(r)\underset{r\rightarrow0}{\simeq}\frac{g}{r^{\alpha}}.~~~~~~~~~~~~~(g>0,~~\alpha<2).
\label{eq:9 V of alpha}
\end{equation}
A series expansion for the reduced radial wave function in the case
of a potential with integer or half-integer $\alpha$ was presented
in \cite{LEZ}. In order to provide calculations of the CP's, e.g.,
for the EP potential with any real $\gamma=-\alpha>-2$, one needs
a series expansion for the corresponding radial wave function. Therefore,
we present such an expansion:
\begin{equation}
\chi(r)=Cr^{\ell+1}\left(1+\sum_{k=1}^{\infty}\sum_{i=0}^{\infty}A_{ki}^{(\ell)}(\beta)r^{k(2-\alpha)+i}\right).\label{eq:10 expansion for chi}
\end{equation}
Here $C$ is an arbitrary constant, and the coefficients $A_{ki}^{(\ell)}(\beta)$
represent polynomials in $\beta$. 
\footnote{However, most of the coefficients are monomials.}
The duplicated powers of $r$ certainly have to be dropped. 
Substituting Eq.\,\eqref{eq:10 expansion for chi} into Eq.\,\eqref{eq:8 Eq.for chi},
and then equating sequentially (starting from the lowest power) the
expansion coefficients of the same powers of $r$ for the left-hand
and right-hand sides, one can calculate
any finite number of the coefficients $A_{ki}^{(\ell)}(\beta)$.
This expansion can be used
to solve Eq.\eqref{eq:8 Eq.for chi}
with or without taking the limit  $E_{n,\ell}\rightarrow 0$ in order to find
either the CP's or the energies $E_{n,\ell}(\beta)$, respectively. 

It is interesting to note that even though the leading term of this
expansion, $r^{\ell+1}$, is well-known, we could not find the general
form \eqref{eq:10 expansion for chi} in the scientific literature.

\section{Cases}


It is easy to check that the Hulthen and the Yukawa potentials fulfill
the condition \eqref{eq:4-}, and therefore their energy levels should
follow \eqref{eq:5inc}. This coincides with the results presented
in \cite{LEZ} (see the first condition in (67) there).

On the contrary, the energy levels of the exponential, the Gaussian
and the Woods-Saxon%
\footnote{Two versions of the Woods-Saxon potential with different parameters
were studied in \cite{LEZ}%
} potentials should satisfy the inequality \eqref{eq:3dec}, according
to the second of conditions (67) in \cite{LEZ}. Interestingly, however,
the Laplacians and the first derivatives of these potentials obey
the conditions
\begin{multline}
\begin{gathered}\Delta V(r)\gtrless0~\textrm{for}~r\lessgtr r_{0},\\
dV(r)/dr>0~\textrm{for all}~r>0
\end{gathered}
~~~~~~~~~~~~~~~~\label{eq:11 Gau cond.}
\end{multline}
which differ slightly from \eqref{eq:2+}, and in fact do not coincide
with any of the conditions mentioned in \cite{BGM}.
Our results for these potentials satisfy inequality \eqref{eq:3dec},
in accord with the conjecture.

Note that the Yukawa and exponential potentials
\footnote{It is worth noting that the potential of the form \eqref{eq:6 the EP V}
was also studied by means of the so called auxiliary field method
in \cite{SSB}. There the critical parameters (or {}``critical heights''
in the language of \cite{SSB}) were obtained for the exponential
and the Yukawa potentials with $\ell,n\leq3$, and within the limits
of their inferior precision, coincide with the results from \cite{LEZ}.}, 
which are the special cases with $\gamma=-1,0$ of the EP potential
\eqref{eq:6 the EP V}, demonstrate opposite ordering for energy levels
with equal principal quantum numbers. The Laplacian of potential \eqref{eq:6 the EP V}
has a form
\footnote{Note that the Laplacian of the Yukawa potential $(\gamma=-1)$ reduces
to the Yukawa potential itself.}:
\begin{equation}
\Delta V(r)=V(r)\left[1+\dfrac{(\gamma+1)(\gamma-2r)}{r^{2}}\right].\label{eq:12 EP Lap}
\end{equation}
It is easily seen that for
$\gamma\leq-1$ the right-hand side of Eq.\,\eqref{eq:12 EP Lap}
is negative for all $r>0$. Hence, the corresponding energy levels
must satisfy inequality~\eqref{eq:5inc}. On the other hand, it follows
from the results of \cite{LEZ} that for $\gamma=0$ 
\footnote{It is easy to guess that for $\gamma>0$, as well.}
the inequality~\eqref{eq:3dec} holds.
This is connected to the fact that the leading term
in expansion of the exponential potential near the origin
is proportional to zero or positive power of $r$
(such is also the case for the Gaussian and Woods-Saxon potentials,
mentioned above).
This behavior suggests that the potential \eqref{eq:6 the EP V}
with $-1<\gamma<0$ may possess some mixed properties.

In order to check this we have computed the CP's for the EP potential
\eqref{eq:6 the EP V} with different $\gamma$ in the range $\left(-1,0\right)$.

\section{A counterexample}

The CP's for EP-potential with $\gamma=-1/2$ are presented in Table
\ref{T1}. One can observe that the critical parameters
$\beta_{n,\ell}$ and $\beta_{n-1,\ell+1}$
(that relate to the same principal quantum number $N$)
are on the diagonals connecting the $\left\lbrace n,\ell\right\rbrace $-positions
assigned as $\left\lbrace n,0\right\rbrace $ on the lower left and
$\left\lbrace 1,n-1\right\rbrace $ on the upper right. The smallest
parameter $\beta_{min}$ of each diagonal was framed.
\footnote{It is seen that the smallest parameters occupy the positions $\left\lbrace n,l\right\rbrace $
with $n=2l+p$, where $p=1,2,3$.}

The critical parameters determine the order in which the energy levels\textit{
appear}.
According to the conjecture formulated in section 2,
it is ``naturally``
assumed that this order is preserved when the parameter $\beta$
of the EP potential~\eqref{eq:6 the EP V} is increased.

If this assumption is true, the energy levels $E_{n,\ell}\left(\beta\right)$
corresponding to $\beta>\beta_{n,\ell}$ must satisfy the inequality
\eqref{eq:5inc} for $\left\lbrace n,\ell\right\rbrace $ belonging
to CP's which are framed or are above the framed ones. On the contrary,
the energy levels for the CP's which are below the framed ones, must
obey the inequality \eqref{eq:3dec}.

The computational results presented in Table \ref{T2} show that this
assumption does not hold here. It follows from Table \ref{T1} that
for the EP potential~\eqref{eq:6 the EP V} with parameters $\gamma=-1/2$
and $\beta>\beta_{1,8}\equiv194.393$ the Schr\"{o}dinger equation \eqref{eq:8 Eq.for chi}
has bound state solutions only for $\ell\leq8$ and $n\leq\left(9-\ell\right)$.
The corresponding binding energies $E_{n,\ell}$ presented in Table
\ref{T2} were computed by two methods for fidelity. First, the results
were obtained by solving Eq.\,\eqref{eq:8 Eq.for chi} directly.
Then, application of the technique presented in Ref.\,\cite{MK}
enabled us to verify the accuracy of these results. Table \ref{T2}
demonstrates that only the four energy levels $E_{n,\ell}(200)$ with
$n+\ell=9~and~5\leq\ell\leq8$ satisfy the inequality~\eqref{eq:5inc},
whereas levels with $\\ 0\leq l \leq 5$ (and the same $N=10$)
fulfill the opposite condition~\eqref{eq:3dec}. This contradicts the assumption, and
disproves the conjecture.

Where does it fail? For the conjecture to hold, the energy levels
must retain the order in which they appear, as
the universal parameter $\beta$
is increased. Obviously, all existing energy levels are lowered when
$\beta$ increases, until a new level appears.
However, for the EP potential \eqref{eq:6 the EP V} with $-1<\gamma<0$
the {}``speed of lowering'' the energy levels $E_{n,\ell}$ and $E_{n',\ell'}$
with $\delta=n-n'=\ell'-\ell=\pm1$ (and the same $N$),
is opposite to the order of their appearance in this process.
That is, a level $E_{n,\ell}$ that has just appeared, will be going
down in energy {}``faster'' than the originally lower level (with
the same $N~and~\delta=\pm1$) that had appeared before it. Therefore, at some value
of $\beta=\beta_{thr}^{(n,\ell)}$ these levels will reverse their order.

\section{Degeneracy and simultaneous appearance}

The fact that the {}``speed of lowering energy'' of the levels that
had just appeared is higher than that of levels with the same
$N~and~\delta=\pm1$ but that had appeared before hand, has several interesting consequences.

First of all, as suggested above, this behavior will cause a \textit{degeneracy}
of two energy levels $E_{n,\ell}=E_{n-1,\ell+1}$, when $\beta$ reaches
the value for which these two levels reverse their order. Indeed,
for the example above (with $\gamma=-1/2$ and $\beta>\beta_{1,8}$),
at some threshold $\beta=\beta_{thr}^{(3,6)}$ the energy $E_{4,5}\left(\beta_{thr}^{(3,6)}\right)$
becomes equal to $E_{3,6}\left(\beta_{thr}^{(3,6)}\right)$. The further
increase of $\beta$ to $\beta_{thr}^{(2,7)}>\beta_{thr}^{(3,6)}$
leads to $E_{2,7}\left(\beta_{thr}^{(2,7)}\right)=E_{3,6}\left(\beta_{thr}^{(2,7)}\right)<E_{4,5}\left(\beta_{thr}^{(2,7)}\right)$.
And at last, for some $\beta=\beta_{thr}^{(1,8)}>\beta_{thr}^{(2,7)}$
one obtains $E_{1,8}\left(\beta_{thr}^{(1,8)}\right)=E_{2,7}\left(\beta_{thr}^{(1,8)}\right)<E_{3,6}\left(\beta_{thr}^{(1,8)}\right)$.
Finally, for $\beta_{thr}^{(1,8)}<\beta<\beta_{1,9}$ all possible
energy levels with $N=10$ satisfy the condition \eqref{eq:3dec}.
The situation is of course similar for smaller $\beta$.
The corresponding $\beta_{thr}$ are presented in Table \ref{T3}, where $\ell_{max}$ is the largest
orbital number which admits a bound state for the given parameter $\beta=\beta_{thr}$.
It is seen that for \u{N}$\equiv\ell_{max}+n=8,9$ there are 3 threshold potentials ($n=1,2,3$) for each \u{N}.
For \u{N}$=5,6,7$ there are only 2 threshold potentials ($n=1,2$) for each \u{N}.
And at last, for \u{N}$=2,3,4$ one obtains the threshold potentials producing
the pairs of states
with $\\ \widetilde{E}_{3p}(\beta_{thr}^{(1,1)})=\widetilde{E}_{3s}(\beta_{thr}^{(1,1)})$, $\widetilde{E}_{4d}(\beta_{thr}^{(1,2)})=\widetilde{E}_{4p}(\beta_{thr}^{(1,2)})$
and $\widetilde{E}_{5f}(\beta_{thr}^{(1,3)})=\widetilde{E}_{5d}(\beta_{thr}^{(1,3)})$
with the same principal quantum numbers.
Here we used the widespread notation for $\widetilde{E}_{N,l}\equiv E_{n,l}$.

The second consequence of this behavior is the possibility for \textit{simultaneous
appearance} of two levels, i.e., there are certain values of $-1<\gamma<0$
for which two new levels $E_{n,\ell}=E_{n-1,\ell+1}(\approx-0)$ {}``appear''
simultaneously at the same value $\beta_{n,\ell}=\beta_{n-1,\ell+1}$.
Our computations show that, e.g.,
\begin{equation}
\beta_{2,0}=\beta_{1,1}\simeq7.9797~~~~\textrm{for}~~~\gamma=-0.238825,\label{14}
\end{equation}

\begin{equation}
\beta_{3,0}=\beta_{2,1}\simeq19.1036~~~~\textrm{for}~~~\gamma=-0.42046,\label{15}
\end{equation}
 and so on.

How does this behavior depend on the parameters of the potential\,\eqref{eq:6 the EP V}?
Recall that the framed values in Table \ref{T1} represent the smallest
$\beta$ for all levels with the same $N~and~\delta=\pm1$. For $\gamma=-1/2$ they
are situated along the ``diagonal'' curve of the table.
As $\gamma$ approaches $0$, the curve
of framed (smallest) $\beta$ approaches the lower left corner. At
the limit $\gamma=0$ all the potentials will satisfy~\eqref{eq:3dec},
regardless of $\beta$. On the contrary, as $\gamma$ approaches $-1$,
the curve of smallest $\beta$ approaches the upper right corner, and
at the limit $\gamma=-1$ all the potentials will satisfy~\eqref{eq:5inc}.

Also note that beyond these limits ($\gamma\leq-1$, or $\gamma\geq0$
) the {}``speed of lowering energy'' of the levels with the same
$N$ and different $\ell$ now \textit{matches} the order of their
appearance. This is related to the behavior of
the potential near the origin. 
As was mentioned above, this is the reason why the exponential,
Gaussian and Woods-Saxon potentials satisfy~\eqref{eq:3dec}, even
though they only fulfill the conditions \eqref{eq:11 Gau cond.} rather
than \eqref{eq:2+}.

\section{Conclusions}

We have used the CPT developed in \cite{LEZ}, together with a new
series expansion for the radial wave function \eqref{eq:10 expansion for chi},
to study the ordering of energy levels with the same principal quantum
number $N$ for a central potential of the form \eqref{eq:6 the EP V},
as well as for several other short-range central potentials widely used
in different fields of physics.
We found that our results are not only consistent with the conditions formulated
in \cite{BGM}, but they even suggest a possible generalization of
the condition \eqref{eq:2+} to \eqref{eq:11 Gau cond.}. We found
that the ordering of levels of the potential \eqref{eq:6 the EP V}
can be characterized according to three ranges of the values of the
parameter $\gamma$
\begin{multline*}
\begin{aligned}E_{n,\ell}>E_{n-1,\ell+1}\qquad\textrm{for}\qquad &~~~~~~~~~ \gamma\geq0\\
E_{n,\ell}\gtreqqless E_{n-1,\ell+1}\qquad\textrm{for}\qquad & -1<\gamma<0\\
E_{n,\ell}<E_{n-1,\ell+1}\qquad\textrm{for}\qquad & -2\leq\gamma\leq-1
\end{aligned}
\end{multline*}

Although it was not shown explicitly, these different properties are
related to the behavior of the potential near the origin, and perhaps
this rule can also be generalized to other potentials.

In the range $-1<\gamma<0$ the above properties give rise to several interesting
consequences for the EP potential \eqref{eq:6 the EP V}, such as the\textit{ degeneracy} 
of two (or more) energy levels
$E_{n,\ell}\left(\beta\right)=E_{n-1,\ell+1}\left(\beta\right)$ at
some calculable values of $\beta=\beta_{thr}$,
or the \textit{simultaneous appearance} of two new energy levels (as
transition states, cf. \cite{LEZ}) at the same value of the critical parameter
$\beta=\beta_{n,\ell}=\beta_{n-1,\ell+1}$.


\section{Acknowledgment}

This work was supported by the Israel Science Foundation grant 954/09.


\begin{table}
\begin{center}
\caption{Critical parameters $\beta_{n,\ell}$ of the EP-potential (\ref{eq:6 the EP V}) with $\gamma=-1/2$, $V(r)=-\beta\exp(-r)/\sqrt{r}$.}
\begin{tabular}{|c||cccccccccc|}
 \hline
 {\large $n \diagdown \ell$}&{\footnotesize 0 }&{\footnotesize 1 }&{\footnotesize 2 }&{\footnotesize 3 }&{\footnotesize 4 }&{\footnotesize 5 }&{\footnotesize 6 }&{\footnotesize 7 }&{\footnotesize 8 }&{\footnotesize 9}
 \tabularnewline
 \hline
 \hline
 {1 }&{\footnotesize 1.72515 }&\multicolumn{1}{|c}{\footnotesize 8.77711 }&{\footnotesize 20.6738 }&{\footnotesize 37.4369 }&{\footnotesize 59.0736 }&{\footnotesize 85.5867 }&{\footnotesize 116.977 }&{\footnotesize 153.246 }&{\footnotesize 194.393 }&{\footnotesize 240.419} \tabularnewline
 \cline{2-2}
 {2 }&{\footnotesize 7.95958 }&\multicolumn{1}{|c}{\footnotesize 19.2665 }&{\footnotesize 35.3988 }&{\footnotesize 56.3915 }&{\footnotesize 82.2555 }&{\footnotesize 112.995 }&{\footnotesize 148.611 }&{\footnotesize 189.105 }&{\footnotesize 234.478 }&{\footnotesize 284.729} \tabularnewline
 \cline{2-3}
 {3 }&{\footnotesize 18.8289 }&\multicolumn{1}{|c|}{\footnotesize 34.2285 }&{\footnotesize 54.5185 }&{\footnotesize 79.6946 }&{\footnotesize 109.755 }&{\footnotesize 144.699 }&{\footnotesize 184.525 }&{\footnotesize 229.232 }&{\footnotesize 278.82 }&{\footnotesize 333.289} \tabularnewline
 \cline{2-3}
 {4 }&{\footnotesize 34.341 }&\multicolumn{1}{|c|}{\footnotesize 53.7313 }&{\footnotesize 78.1078 }&{\footnotesize 107.418 }&{\footnotesize 141.639 }&{\footnotesize 180.761 }&{\footnotesize 224.775 }&{\footnotesize 273.678 }&{\footnotesize 327.468 }&{\footnotesize 386.142} \tabularnewline
 \cline{3-4}
 {5 }&{\footnotesize 54.498 }&\multicolumn{1}{|c|}{\footnotesize 77.8093 }&\multicolumn{1}{c|}{\footnotesize 106.213 }&{\footnotesize 139.61 }&{\footnotesize 177.956 }&{\footnotesize 221.226 }&{\footnotesize 269.405 }&{\footnotesize 322.485 }&{\footnotesize 380.46 }&{\footnotesize 443.325} \tabularnewline
 \cline{3-4}
 {6 }&{\footnotesize 79.3008 }&{\footnotesize 106.482 }&\multicolumn{1}{|c|}{\footnotesize 138.864 }&{\footnotesize 176.307 }&{\footnotesize 218.742 }&{\footnotesize 266.131 }&{\footnotesize 318.451 }&{\footnotesize 375.687 }&{\footnotesize 437.829 }&{\footnotesize 504.872} \tabularnewline
 \cline{4-5}
 {7 }&{\footnotesize 108.75 }&{\footnotesize 139.762 }&\multicolumn{1}{|c}{\footnotesize 176.081 }&\multicolumn{1}{|c|}{\footnotesize 217.532 }&{\footnotesize 264.024 }&{\footnotesize 315.505 }&{\footnotesize 371.942 }&{\footnotesize 433.313 }&{\footnotesize 499.606 }&{\footnotesize 570.809} \tabularnewline
 \cline{4-5}
 {8 }&{\footnotesize 142.845 }&{\footnotesize 177.657 }&{\footnotesize 217.879 }&\multicolumn{1}{|c|}{\footnotesize 263.306 }&{\footnotesize 313.825 }&{\footnotesize 369.371 }&{\footnotesize 429.901 }&{\footnotesize 495.388 }&{\footnotesize 565.812 }&{\footnotesize 641.161} \tabularnewline
 \cline{5-6}
 {9 }&{\footnotesize 181.588 }&{\footnotesize 220.174 }&{\footnotesize 264.269 }&\multicolumn{1}{|c|}{\footnotesize 313.642 }&\multicolumn{1}{|c|}{\footnotesize 368.162 }&{\footnotesize 427.749 }&{\footnotesize 492.35 }&{\footnotesize 561.931 }&{\footnotesize 636.47 }&{\footnotesize 715.947} \tabularnewline
 \cline{5-6}
 {10 }&{\footnotesize 224.977 }&{\footnotesize 267.316 }&{\footnotesize 315.259 }&{\footnotesize 368.553 }&\multicolumn{1}{|c|}{\footnotesize 427.049 }&{\footnotesize 490.653 }&{\footnotesize 559.304 }&{\footnotesize 632.961 }&{\footnotesize 711.596 }&{\footnotesize 795.187} \tabularnewline
 \cline{6-7}
 {11 }&{\footnotesize 273.013 }&{\footnotesize 319.087 }&{\footnotesize 370.857 }&{\footnotesize 428.048 }&\multicolumn{1}{|c|}{\footnotesize 490.497 }&\multicolumn{1}{c|}{\footnotesize 558.097 }&{\footnotesize 630.778 }&{\footnotesize 708.493 }&{\footnotesize 791.207 }&{\footnotesize 878.895} \tabularnewline
 \cline{6-7}
 {12 }&{\footnotesize 325.696 }&{\footnotesize 375.489 }&{\footnotesize 431.066 }&{\footnotesize 492.134 }&{\footnotesize 558.515 }&\multicolumn{1}{|c|}{\footnotesize 630.091 }&{\footnotesize 706.784 }&{\footnotesize 788.538 }&{\footnotesize 875.315 }&{\footnotesize 967.086} \tabularnewline
  \cline{7-8}
 {13 }&{\footnotesize 383.027 }&{\footnotesize 436.525 }&{\footnotesize 495.891 }&{\footnotesize 560.818 }&{\footnotesize 631.112 }&\multicolumn{1}{|c|}{\footnotesize 706.646 }&\multicolumn{1}{c|}{\footnotesize 787.332 }&{\footnotesize 873.11 }&{\footnotesize 963.934 }&{\footnotesize 1059.77} \tabularnewline
  \cline{7-8}
 {14 }&{\footnotesize 445.004 }&{\footnotesize 502.195 }&{\footnotesize 565.336 }&{\footnotesize 634.104 }&{\footnotesize 708.294 }&{\footnotesize 787.769 }&\multicolumn{1}{|c|}{\footnotesize 872.432 }&{\footnotesize 962.217 }&{\footnotesize 1057.07 }&{\footnotesize 1156.97} \tabularnewline
  \cline{8-9}
 {15 }&{\footnotesize 511.629 }&{\footnotesize 572.501 }&{\footnotesize 639.403 }&{\footnotesize 711.998 }&{\footnotesize 790.068 }&{\footnotesize 873.467 }&\multicolumn{1}{|c|}{\footnotesize 962.091 }&\multicolumn{1}{c|}{\footnotesize 1055.87 }&{\footnotesize 1154.74 }&{\footnotesize 1258.67} \tabularnewline
 \cline{8-9}
 {16 }&{\footnotesize 582.901 }&{\footnotesize 647.445 }&{\footnotesize 718.096 }&{\footnotesize 794.502 }&{\footnotesize 876.437 }&{\footnotesize 963.745 }&{\footnotesize 1056.32 }&\multicolumn{1}{|c|}{\footnotesize 1154.07 }&{\footnotesize 1256.95 }&{\footnotesize 1364.91} \tabularnewline
 \cline{9-10}
 {17 }&{\footnotesize 658.82 }&{\footnotesize 727.027 }&{\footnotesize 801.414 }&{\footnotesize 881.62 }&{\footnotesize 967.407 }&{\footnotesize 1058.61 }&{\footnotesize 1155.11 }&\multicolumn{1}{|c|}{\footnotesize 1256.83 }&\multicolumn{1}{c|}{\footnotesize 1363.7 }&{\footnotesize 1475.68} \tabularnewline
 \cline{9-10}
 {18 }&{\footnotesize 739.386 }&{\footnotesize 811.249 }&{\footnotesize 889.362 }&{\footnotesize 973.355 }&{\footnotesize 1062.98 }&{\footnotesize 1158.07 }&{\footnotesize 1258.49 }&{\footnotesize 1364.16 }&\multicolumn{1}{|c|}{\footnotesize 1475.01 }&{\footnotesize 1590.98} \tabularnewline
 \cline{10-10}
 {19 }&{\footnotesize 824.6 }&{\footnotesize 900.11 }&{\footnotesize 981.94 }&{\footnotesize 1069.71 }&{\footnotesize 1163.16 }&{\footnotesize 1262.12 }&{\footnotesize 1366.45 }&{\footnotesize 1476.06 }&\multicolumn{1}{|c|}{\footnotesize 1590.87 }&{\footnotesize 1710.84} \tabularnewline
 \cline{10-10}
 {20 }&{\footnotesize 914.46 }&{\footnotesize 993.611 }&{\footnotesize 1079.15 }&{\footnotesize 1170.68 }&{\footnotesize 1267.95 }&{\footnotesize 1370.77 }&{\footnotesize 1478.99 }&{\footnotesize 1592.53 }&{\footnotesize 1711.3 }&{\footnotesize 1835.24} \tabularnewline
  \hline
\end{tabular}
\label{T1}
\end{center}
\end{table}

\begin{table}
\begin{center}
\caption{Binding energies $-E_{n,\ell}$ for the EP-potential $V(r)=-200\exp(-r)/\sqrt{r}$.}
\begin{tabular}{|c|ccccccccc}
\hline
{\large $n\diagdown \ell$} & {\small 0} & {\small 1} & {\small 2} & {\small 3} & {\small 4} & {\small 5} & {\small 6} & {\small 7} & \multicolumn{1}{c|}{{\small 8}}\tabularnewline
\hline
{\small 1} & {\footnotesize 453.076} & {\footnotesize 253.517} & {\footnotesize 161.668} & {\footnotesize 106.560} & {\footnotesize 69.5605} & {\footnotesize 43.3235} & {\footnotesize 24.2864} & {\footnotesize 10.5122} & \multicolumn{1}{c|}{{\footnotesize 0.943082}}\tabularnewline
\cline{10-10}
{\small 2} & {\footnotesize 217.692} & {\footnotesize 142.656} & {\footnotesize 95.1156} & {\footnotesize 62.4832} & {\footnotesize 39.1216} & {\footnotesize 22.1330} & {\footnotesize 9.89629} & \multicolumn{1}{c|}{{\footnotesize 1.52600}} & \multicolumn{1}{c}{}\tabularnewline
\cline{9-9}
{\small 3} & {\footnotesize 123.435} & {\footnotesize 83.5646} & {\footnotesize 55.2885} & {\footnotesize 34.7710} & {\footnotesize 19.7922} & {\footnotesize 9.04198} & \multicolumn{1}{c|}{{\footnotesize 1.79035}} & \multicolumn{1}{c}{} & \multicolumn{1}{c}{}\tabularnewline
\cline{8-8}
{\small 4} & {\footnotesize 71.9016} & {\footnotesize 48.0047} & {\footnotesize 30.3073} & {\footnotesize 17.3037} & {\footnotesize 7.99882} & \multicolumn{1}{c|}{{\footnotesize 1.81080}} & \multicolumn{1}{c}{} &  & \multicolumn{1}{c}{}\tabularnewline
\cline{7-7}
{\small 5} & {\footnotesize 40.6508} & {\footnotesize 25.7661} & {\footnotesize 14.7085} & {\footnotesize 6.81296} & \multicolumn{1}{c|}{{\footnotesize 1.64754}} & \multicolumn{1}{c}{} &  &  & \multicolumn{1}{c}{}\tabularnewline
\cline{6-6}
{\small 6} & {\footnotesize 21.1813} & {\footnotesize 12.0487} & {\footnotesize 5.53120} & \multicolumn{1}{c|}{{\footnotesize 1.35410}} & \multicolumn{1}{c}{} &  &  &  & \multicolumn{1}{c}{}\tabularnewline
\cline{5-5}
{\small 7} & {\footnotesize 9.36962} & {\footnotesize 4.20250} & \multicolumn{1}{c|}{{\footnotesize 0.982547}} & \multicolumn{1}{c}{} &  &  &  &  & \multicolumn{1}{c}{}\tabularnewline
\cline{4-4}
{\small 8} & {\footnotesize 2.88204} & \multicolumn{1}{c|}{{\footnotesize 0.58778}} & \multicolumn{1}{c}{} &  &  &  &  &  & \tabularnewline
\cline{3-3}
\multicolumn{1}{|c|}{{\small 9}} & \multicolumn{1}{c|}{{\footnotesize 0.234378}} & \multicolumn{1}{c}{} &  &  &  &  &  &  & \tabularnewline
\cline{1-1} \cline{2-2}
\end{tabular}
\label{T2}
\end{center}
\end{table}

\begin{table}
\begin{center}
\caption{Threshold parameters $\beta_{thr}$ of EP-potential with $\gamma=-1/2$ for $E_{(n,\ell_{max})}(\beta_{thr})=E_{(n+1,\ell_{max}-1)}(\beta_{thr})$.}
\begin{tabular}{|c||c|c|c|c|c|c|c|c|}
\hline
{\large$n\diagdown(\ell_{max}+n)$} & {\small 2} & {\small 3} & {\small 4} & {\small 5} & {\small 6} & {\small 7} & {\small 8} & {\small 9}\tabularnewline
\hline
\hline
{\small 1} & {\footnotesize 9.002821} & {\footnotesize 22.22577} & {\footnotesize 41.3813} & {\footnotesize 66.4516} & {\footnotesize 97.4325} & {\footnotesize 134.3227} & {\footnotesize 177.1216} & {\footnotesize 225.82855}\tabularnewline
\hline
{\small 2} &  &  &  & {\footnotesize 59.9317} & {\footnotesize 89.30232} & {\footnotesize 124.6208} & {\footnotesize 165.8688} & {\footnotesize 213.0374}\tabularnewline
\hline
{\small 3} &  &  &  &  &  &  & {\footnotesize 155.5733} & {\footnotesize 201.1049}\tabularnewline
\hline
\end{tabular}
\label{T3}
\end{center}
\end{table}


\begin{thebibliography}{99}

\bibitem{BGM} B. Baumgartner, H. Grosse, and A. Martin, \emph{The
Laplacian of the potential and the order of energy levels}, Phys.
Lett. B 146 (1984) 363-366.

\bibitem{GM} H. Grosse and A. Martin, \emph{Two theorems on the level
order in potential model}, Phys. Lett. B 134 (1984) 368-372.

\bibitem{FFD} G. Feldman, T. Fulton, and A. Devoto, \emph{Energy
levels and level ordering in the WKB approximation}, Nucl. Phys. B
154 (1979) 441-462.

\bibitem{GMS} H. Grosse, A. Martin, and J. Stubbe, \emph{Order of
energy levels for the Klein-Gordon equation}, Phys. Lett. B 255 (1991)
563-566.

\bibitem{LEZ} E. Z. Liverts and N. Barnea, \emph{Transition states
and the critical parameters of central potentials}, J. Phys. A: Math.
Theor. 44 (2011) 375303 (24 pp).

\bibitem{SSB} B. Silvestre-Brac, C. Semay and F. Buisseret, \textit{The
auxiliary field method and approximate analytical solutions of the
Schr\"{o}dinger equation with exponential potentials}, J. Phys. A: Math.
Theor. 42 (2009) 245301 (18 pp)

\bibitem{WM} http://www.wolfram.com/mathematica/

\bibitem{MK} R. Krivec and V. B. Mandelzweig, \textit{Quasilinearization
approach to quantum mechanics}, Comp. Phys. Commun. 152 (2003) 165-174.


\end{thebibliography}
\end{document}